\documentclass[11pt]{elsart}
\usepackage[dvips]{graphicx}

\begin{document}
\begin{frontmatter}

\title{High dimensional random Apollonian networks}

\author{Zhongzhi Zhang\qquad}
\ead{xinjizzz@sina.com}
\author{Lili Rong}
\address{
Institute of System Engineering, Dalian University of Technology\\
2 Ling Gong Rd., Dalian 116023, Liaoning, China}%
\ead{llrong@dlut.edu.cn}
\author{Francesc Comellas\corauthref{fc}}
\corauth[fc]{Corresponding author. Tel. +34-93-4137072.}
\address{
Dep. de Matem\`atica Aplicada IV, EPSC, Universitat
Polit\`ecnica de Catalunya\\
 Av. Canal Ol\'{\i}mpic s/n, 08860
Castelldefels, Barcelona, Catalonia, Spain}
\ead{comellas@mat.upc.es}

\begin{abstract}
We propose a simple algorithm which produces a new category of
networks, high dimensional random Apollonian networks, with
small-world and scale-free characteristics. We derive analytical
expressions for their degree distributions and clustering
coefficients which are determined by the dimension of the network.
The values obtained for these parameters are in good agreement
with simulation results and comparable to those coming from real
networks. We estimate also analytically that the average path length
of the networks increases at most logarithmically with the number
of vertices.
\end{abstract}
%\pacs{89.20.Hh, 89.75.Hc, 89.75.Da}
%89.20.Hh World Wide Web, Internet
%89.75.Da Systems obeying scaling laws
%89.75.Fb Structures and organization in complex systems
%89.75.-k Complex systems
%89.75.Hc Networks and genealogical trees

\date{}
\end{frontmatter}

%%%%%%%%%%%%%%%%%%%%%%%%%%%%%%%%%%%%%%%%%%%%%%%%%%%%%%%%%%%%%%%%%
%%%%%%%%%%%%%%%%%%%%%%%%%%%%%%%%%%%%%%%%%%%%%%%%%%%%%%%%%%%%%%%%%
%\vskip -0.5cm\color{Blue}
%\vbox to 0pt{\kern -14cm {
%\noindent \small \copyright 2005
%{\em Elsevier Science B.V. All rights reserved}\\
%{\em Physica A}, submitted.}
%\vss}\color{Black}

%%%%%%%%%%%%%%%%%%%%%%%%%%%%%%%%%%%%%%%%%%%%%%%%%%%%%%%%%%%%%%%%%%%%
\section{Introduction}
Since the pioneering papers by Watts and Strogatz on small-world
networks ~\cite{WaSt98} and Barab\'asi and Albert on scale-free
networks~\cite{BaAl99}, complex networks have received considerable
attention as an interdisciplinary subject~\cite{AlBa02,DoMe02,Ne03}.
Complex networks describe many systems in nature and society, such
as Internet~\cite{FaFaFa99}, World Wide Web ~\cite{AlJeBa99},
metabolic networks ~\cite{JeToAlOlBa00}, protein networks in the
cell ~\cite{JeMaBaOl01}, co-author networks ~\cite{Ne01} and
sexual networks ~\cite{LiEdAmStAb01}, most of which share three
apparent features: power-law degree distribution, small average
path length (APL) and high clustering coefficient.
  In recent years, many evolving models~\cite{AlBa02,DoMe02,Ne03}
have been proposed to describe real-life networks. The original BA
model ~\cite{BaAl99,BaRaVi01} captures the two main mechanisms
responsible for the power-law degree distribution of degrees,
namely growth and preferential attachment. Dorogovtsev, Mendes,
and Samukhin ~\cite{DoMeSa00} gave an exact solution for a class
of growing network models thanks to the use of a
``master-equation''. Krapivsky, Redner, and Leyvraz
\cite{KaReLe00} examined the effect of a nonlinear preferential
attachment on network dynamics and topology. Amaral et al.
\cite{AmScBaSt00} studied models that incorporate aging and cost
and capacity constraints in order to explain the deviations from
the power-law behavior in several real-life networks. Dorogovtsev
and Mendes \cite{DoMe00} also addressed the evolution of networks
with aging of sites. Bianconi and Barab\'asi \cite{BiBa01} offered
a model addressing the competitive aspect in many real networks
such as World Wide Web. Additionally, in real systems a series of
microscopic events shape the network evolution, including the
addition or rewiring of new edges or the removal of vertices or
edges. Albert and Barab\'asi \cite{AlBa00} discussed a model that
incorporates new edges between existing vertices and the rewiring
of edges. Dorogovtsev and Mendes \cite{DoMe00b} considered a class
of undirected models in which new edges are added between old
vertices and existing edges can be removed. Although it is now
established that preferential attachment can explain the power-law
characteristic of networks, there is a wide range of microscopic
alternative mechanisms that could affect the evolution of growing
networks and still lead to the observed scale-free topologies.
Kleinberg et al. \cite{KlKuRaRaTo99} and Kumar et al.
\cite{KuRaRaSiToUp00,KuRaRaSiToUp00b} proposed copying mechanisms
motivated by the desire to explain the power-law degree
distribution of the World Wide Web. Chung et al.
~\cite{ChLuDeGa03} introduced also a duplication model for
biological networks. Krapivsky and Render \cite{KrRe01} presented
edge redirection mechanisms which are mathematically equivalent to
the model  of Kumar et al \cite{KuRaRaSiToUp00,KuRaRaSiToUp00b}.
Inspired by citation networks V\'azquez proposed in  ~\cite{Va01}
the walking mechanism. Actually scale-free networks can be created
by various methods. Comellas, Fertin and Raspaud \cite{CoFeRa04}
introduced a category of graphs via a recursive construction.
Specific recursive and general scale-free constructions are given also
in~\cite{BaRaVi01,CoSa02,DoGoMe02,RaBa03,No03} and scale-free
trees (without clustering) in~\cite{JuKiKa02}. Zhou et al. construct in ~\cite{ZhWaHuCh04} an integer network using a pure mathematical method.

%In relation to the problem of Apollonian packing, see Fig. 1,  
%two groups independently proposed the Apollonian networks 
%\cite{AnHeAnSi05,DoMa05}.

In relation to the problem of Apollonian packing, a two-dimensional example
of which is shown in Fig. 1, Andrade et al. introduced Apollonian
networks~\cite{AnHeAnSi05} which were also proposed by
Doye and Massen in~\cite{DoMa05}. 
It should be pointed out that high-dimensional Apollonian networks
were already introduced in \cite{DoMa05}, but in this work
the emphasis is placed on two-dimensional Apollonian networks and 
to provide a model to help understanding the energy landscape networks.
Zhang et al. in \cite{ZhCoFeRo05} offer a simple general algorithm
producing high-dimensional Apollonian networks and derive in detail
analytical expressions for their order, size, degree distribution, clustering
coefficient and  diameter. Inspired also by the Apollonian packing,
Zhou et al. \cite{ZhYaZhFuWa04} proposed a simple rule that
generates random two-dimensional Apollonian networks with very large clustering
coefficient and very small APL.
%%%%%%%%%%%%%%%%%%%%%%%%%%%%%%%%%%%%%%%%%%%%%%%%%%%%%%%%%%
% Figure  1
%%%%%%%%%%%%%%%%%%%%%%%%%%%%%%%%%%%%%%%%%%%%%%%%%%%%%%%%%%
\begin{figure}
\begin{center}
\includegraphics[width=6cm]{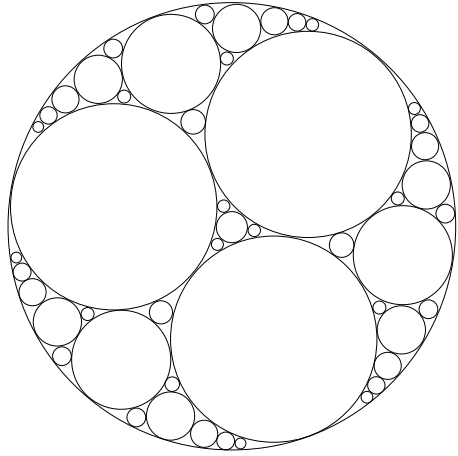}
\caption{An Apollonian packing of disks within a circle.}
\label{fig:Apollo}
\end{center}
\end{figure}
%%%%%%%%%%%%%%%%%%%%%%%%%%%%%%%%%%%%%%%%%%%%%%%%%%%%%%%%%%

In this paper, motivated by  high-dimensional Apollonian
packings, we present a simple iterative algorithm for high-dimensional random Apollonian networks (HDRAN) which extends the
idea introduced in \cite{ZhYaZhFuWa04} for dimension two to any
dimension. The algorithm can concretize the problems of abstract 
high-dimensional random Apollonian packings. 
Using the algorithm we determine relevant
characteristics of HDRAN: scale-free degree distribution and large
clustering coefficient which depend on the dimension of the
Apollonian packing. Based on the algorithm, we estimate also that
the average path length of the networks increases logarithmically
with the number of vertices. Therefore, these networks (HDRAN)
have a small-world scale-free topology.

%%%%%%%%%%%%%%%%%%%%%%%%%%%%%%%%%%%%%%%%%%%%%%%%%%%%%%%%%%%%%%%%%%%
%%%%%%%%%%%%%%%%%%%%%%%%%%%%%%%%%%%%%%%%%%%%%%%%%%%%%%%%%%%%%%%%%%%
%%%%%%%%%%%%%%%%%%%%%%%%%%%%%%%%%%%%%%%%%%%%%%%%%%%%%%%%%%%%%%%%%%%
\section{The construction of high-dimensional random Apollonian networks}
The name of Apollonian packings goes back to Apollonius of Perga 
(c. 262 BC-- c. 190 BC) who studied the problem of finding 
tangent circles to any three given circles.
When the initial three circles are mutually tangent, 
there exist only two of such tangent circles, known as 
Apollonian circles or Soddy inner and outer circles. 
Frederick Soddy~\cite{So36}, Nobel Prize for Chemistry in 1921,  published in Nature 
a poem describing the equation which relates the radii of the circles and
is now known as  the ``Kissing Circles Theorem''.  
This result, however, had been previously given by Ren\'e Descartes in 1643 and
rediscovered in 1842 by the mathematician Phillip Beecroftan.
A generalization to $n$-dimensional hyperspheres, also as a poem, 
was published as well in Nature by Thorold Gosset~\cite{Go37} and reads:
\begin{quotation}
\em\small  
And let us not confine our cares\\
To simple circles, planes and spheres,\\
But rise to hyper flats and bends\\
Where kissing multiple appears,\\
In $n$-ic space the kissing pairs\\
Are hyperspheres, and Truth declares -\\
As $n+2$ such osculate\\
Each with an $n+1$ fold mate\\
The square of the sum of all the bends\\
Is $n$ times the sum of their squares.
\end{quotation}
In the procedure known as {\em Apollonian packing} or  {\em Apollonian gasket}, 
we start with three mutually tangent circles, and draw their inner Soddy circles. 
Then we draw the inner Soddy circles of this circle with each pair of the original three, 
and the process is iterated, see Fig.1.
This Apollonian packing can be used to design a network, where
each circle is associated to a vertex of the network and vertices 
are connected if the corresponding circles are tangent.
We call this network a two-dimensional Apollonian network~\cite{AnHeAnSi05,DoMa05}.

In a similar way, it is possible to construct higher dimensional
($d$-dimensional $d\geq 2$) Apollonian networks associated
with other self-similar packings.
We start with $d+1$ mutually tangent $d$-dimensional hyperspheres
($d$-hyperspheres) in a $d$-dimensional simplex
which is enclosed  and tangent to a larger hypersphere, 
the interstices of which are curvilinear $d$-dimensional simplices ($d$-simplices).
In the first iteration, and inside each of the interstices, we add
$d$-hyperspheres tangent to each of the $d+1$ $d$-hyperspheres bounding the curvilinear 
$d$-simplex. The added hyperspheres will not fill the interstice and 
each produces $d+1$ smaller interstices. Then, in the second iteration, $d+1$ 
$d$-hyperspheres are added inside all of the $d+1$ new interstices, being 
again tangent to the enclosing $d$-hyperspheres.
The process is repeated iteratively obtaining 
the high-dimensional Apollonian packing.
If each $d$-hypersphere corresponds to a vertex and vertices
are connected by an edge if the corresponding $d$-hyperspheres
are tangent, then we obtain a $d$-dimensional Apollonian network.

  In the above $d$-dimensional Apollonian packings, if we add
only one $d$-hypersphere inside a randomly selected interstice
in every iteration, after a certain number of iterations we have a high-dimensional
random Apollonian packing. Analogously, if each $d$-hypersphere
corresponds to a vertex and vertices are connected by an edge
if the corresponding $d$-hyperspheres are tangent, then one
obtains $d$-dimensional random Apollonian networks.
In \cite{ZhYaZhFuWa04} it is discussed the special case  $d=2$.
Here we will consider the general case.
It should be noted that we consider the initial bounding
$d$-hypersphere at the same level as the other
$d+1$ initial $d$-hyperspheres.

%%%%%%%%%%%%%%%%%%%%%%%%%%%%%%%%%%%%%%%%%%%%%%%%%%%%%%%%%%%%%%%%
%%%%%%%%%%%%%%%%%%%%%%%%%%%%%%%%%%%%%%%%%%%%%%%%%%%%%%%%%%%%%%%%
\section{The iterative algorithm for high-dimensional random Apollonian networks}
In the iterative process for the construction of HDRAN,
at each step and for each new hypersphere added, $d+1$ new interstices
are created in the packing,  that will be filled in one of the following iterations.
Equivalently, for each new vertex added, $d+1$ new  $d$-simplices
are created in the network, into which vertices will be inserted in one
of the following iterations.
According to this process, we introduce a general iterative
algorithm for HDRAN.
We denote the $d$-dimensional random Apollonian network after $t$ iterations
by $A(d,t)$, $d\geq 2$.
Before introducing the algorithm we give the following definition.
A complete graph $K_d$ (also referred in the literature as $d$-clique; see \cite{We01})
is a graph without loops whose $d$ vertices are pairwise adjacent.
Then the $d$-dimensional random Apollonian network $A(d,t)$
is constructed as follows.
Initially ($t=0$), $A(d,0)$ is the complete graph $K_{d+2}$ or $(d+2)$-clique.
At each step, we choose an existing subgraph isomorphic to a
$(d+1)$-clique {\em that has never been selected before}, then we add a new vertex
and join it to all the vertices of the selected subgraph.
The growing process is repeated until the network reaches the desired size.
Note that when $d=2$, our model gives the maximal planar
network of \cite{ZhYaWa04}.
Furthermore, since the network size is incremented by one with each
step, in this paper, we use the step value $t$ to represent a
vertex created at this step.
We can see easily that at step $t$, the network consists of $N=t+d+2$ vertices.
The total degree equals $(d+1)(2t+d+2)$. So, when $t$ is large the average vertex
degree at step $t$ is equal approximately to a constant value $2(d+1)$, which shows
that the network $A(d,t)$ is sparse like many real-life
networks ~\cite{AlBa02,DoMe02,Ne03}.

%%%%%%%%%%%%%%%%%%%%%%%%%%%%%%%%%%%%%%%%%%%%%%%%%%%%%%%%%%%%%%
%%%%%%%%%%%%%%%%%%%%%%%%%%%%%%%%%%%%%%%%%%%%%%%%%%%%%%%%%%%%%%%%
\section{Relevant characteristics of high-dimensional random Apollonian networks}
In this section we show that the dimension $d$ is a tunable parameter which
controls the relevant characteristics of a high dimensional random
Apollonian network $A(d,t)$.

\subsection{Degree distribution}
The degree distribution is one of the most important statistical
characteristics of a network.
We use the mean-field method~\cite{AlBa02,BaAlJe99} to predict the growth
dynamics of the individual vertices, and calculate analytically
the connectivity distribution $P(k)$ and the scaling exponents.
Note that, after a new vertex is added, the number of $(d+1)$-cliques that
can be chosen in the following step increases by $d$.
Therefore, we can immediately see that after $t$ steps, the number of
$(d+1)$-cliques ready for selection are $d t+d+2$.
Given a vertex, when its degree increases by one, the number of $(d+1)$-cliques
that contain the vertex increases by $d-1$, so the number of $(d+1)$-cliques for selection
containing the vertex $i$ with degree $k_i$ is $(d-1)k_i-d^2+d+2$.
If we consider $k$ continuous, we can write for a vertex $i$
%%%
%%% eq 1
\begin{equation}
{\partial k_i \over \partial t} ={(d-1)k_i-d^2+d+2 \over dt+d+2}
\end{equation}
The solution of this equation, with the initial condition that vertex $i$
was added to the network at  $t_i$ with connectivity $k_i(t_i)=d+1$, is
%%%
%%% eq 2
\begin{equation}
k_i(t) ={d^2-d-2 \over d-1}+{d+1\over d-1}\left({dt+d+2 \over dt_i+d+2}\right)^{d-1 \over d}
\end{equation}
The probability that a vertex has a connectivity $k_i(t)$ smaller
than $k$, $P(k_i(t)<k)$, is
%%%
%%% eq 3
\begin{equation}
P(k_i(t)<k) =P\left(t_i>{{(dt+d+2)\left({d+1 \over d-1}\right)^{d \over d-1}} \over
  {d\left(k-{d^2-d-2\over d-1}\right)^{d\over d-1}}}-{d+2\over d}\right)
\end{equation}
Assuming that we add the vertices to the network  at equal intervals,
the probability density of $t_i$ is
%%%
%%% eq 4
\begin{equation}
P_i(t_i) ={1\over d+2+t}
\end{equation}
Substituting this into Eq. (3) we obtain that
%%%
%%% eq 5
\begin{eqnarray}
\lefteqn{P\left(t_i > {{(dt+d+2)\left({d+1\over d-1}\right)^{d \over d-1}} \over
  {d\left(k-{d^2-d-2\over d-1}\right)^{d\over d-1}}}-{d+2\over d}\right)  = }\\
&&= 1-P\left(t_i\leq {{(dt+d+2)\left({d+1\over d-1}\right)^{d \over d-1}} \over
  {d\left(k-{d^2-d-2\over d-1}\right)^{d\over d-1}}}-{d+2\over d}\right)= \nonumber \\
&&=  1-{{(dt+d+2)\left({d+1\over d-1}\right)^{d \over d-1}} \over
  {(d+2+t)d\left(k-{d^2-d-2\over d-1}\right)^{d\over d-1}}}+{d+2\over (d+2+t)d} \nonumber
\end{eqnarray}

Thus the degree distribution is
%%%
%%% eq 6
\begin{equation}
P(k)={\partial P(k_i(t)<k)\over \partial k} =
{(dt+d+2)(d+1)^{d \over d-1}\over (d+2+t)}
\left((d-1)k-{(d^2-d-2)}\right)^{1-2d \over d-1}
\end{equation}

For large $t$
%%%
%%% eq 7
\begin{equation}
P(k)= {d(d+1)^{d \over d-1}}
\left((d-1)k-{(d^2-d-2)}\right)^{1-2d \over d-1}
\end{equation}
and if $k \gg d$ then $P(k)\sim  k^{-\gamma}$  giving a degree exponent 
$\gamma(d)={2d-1 \over d-1}$.

This value is different from the degree exponent for a high dimensional 
Apollonian network (non random HDAN) which is $\gamma=1+ {{\log (d+1)}\over{\log d}}$
~\cite{DoMa05,ZhCoFeRo05}. 
This discrepancy would need further study, but a possible explanation is that 
for two networks with the same number of vertices, HDRAN and HDAN,  the 
distribution of degrees differs because of the different growing process.
The method used to obtain $P(k)$ might have also some influence in the value of 
$\gamma$ as it is explained at the end of the next subsection.

When $d=2$, one has $\gamma(2)=3$~\cite{ZhYaZhFuWa04},
while as $d$ goes to infinity $\gamma(\infty)=2$.
Thus by tuning the parameter $d$, it is possible to obtain
a variety of scale-free networks
with different exponents in the range, $2<\gamma<3$.
In Fig. 2, the degree distributions $P(k)$ at various values of the dimension $d$
are displayed and we find that at any value of $d$, the simulated degree
distribution is in good agreement with the analytic value.

%%%%%%%%%%%%%%%%%%%%%%%%%%%%%%%%%%%%%%%%%%%%%%%%%%%%%%%%%%
% Figure  2
%%%%%%%%%%%%%%%%%%%%%%%%%%%%%%%%%%%%%%%%%%%%%%%%%%%%%%%%%%
\begin{figure}
\begin{center}
\includegraphics[width=12cm]{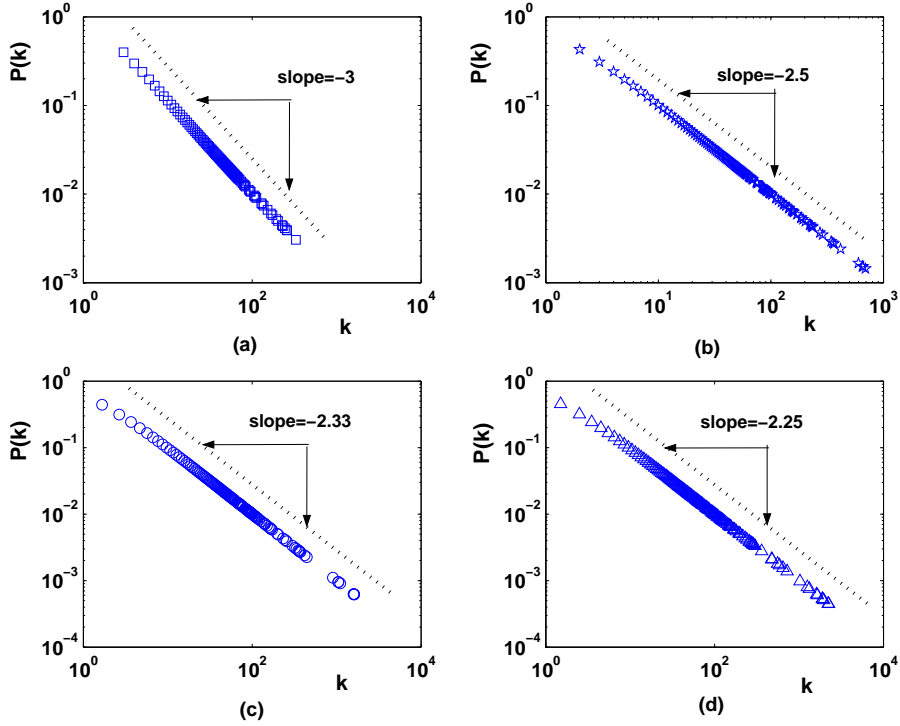}
\caption{Log-log graph of the cumulative degree distribution of HDRAN with
order $N=10000$ at various values of the dimension $d$. The squares
(a), stars (b),  circles (c), and triangles (d) denote the
simulation results for $d=2$, $d=3$, $d=4$, and $d=5$,
respectively. The reference dotted lines come from the evaluation
of  $\gamma(d)$ as given by  equation (7).}
\label{fig:loglog}
\end{center}
\end{figure}
%%%%%%%%%%%%%%%%%%%%%%%%%%%%%%%%%%%%%%%%%%%%%%%%%%%%%%%%%%

Also, from Eqs. (2) and (7), we can notice that when $t$ becomes large,
the maximal degree of a vertex is roughly $N^{1/(\gamma-1)}$.

\subsection{Clustering coefficient}
The clustering coefficient of a vertex is the ratio of the total
number of existing connections between all its $k$ nearest neighbors
and $k(k-1)/2$, the number of all possible connections between them.
The clustering of the graph is obtained averaging over all its vertices.
We can derive analytical expressions for the clustering $C(k)$ for any
vertex with degree $k$.
When a vertex is created it is connected to all the vertices of a
$(d+1)$-clique whose vertices are completely interconnected.
It follows that a vertex with degree $k=d+1$ has a clustering coefficient of one because all the $(d+1)d/2$ possible links between its neighbors actually exist.
In the following steps,  a newly-created vertex links
the considered vertex whose $d$ nearest neighbors are also connected
by the new one at the same iteration step. Thus there is a one-to-one correspondence between the degree of a vertex and its clustering.
For a vertex $v$ of degree $k$, the exact expression for its clustering
coefficient is
%%%
%%% eq 8
\begin{equation}
C(k)= {{{d(d+1)\over 2}+ d(k-d-1)} \over {k(k-1)\over 2}}= {d(2k-d-1)\over k(k-1)}
\end{equation}
which depends on the degree $k$ and the dimension $d$.
This expression indicates that the local clustering scales as
$C(k)\sim k^{-1}$ for large $k$. Note that, as the clustering
only depends on $k$ for a given dimension, it matches the
results found in ~\cite{DoMa05} for HDAN.
It is interesting to notice that a similar scaling
has been observed in several real-life networks~\cite{RaBa03}.

The clustering coefficient $C$ of HDRAN can be obtained as the mean value
of $C(k)$ with respect to the degree distribution $P(k)$ expressed by Eq. (7).
The result is
%%%
%%% eq 9
\begin{eqnarray}
C& = &\int_{d+1}^{\infty}C(k)P(k)\, dk =\nonumber \\
    & = & \int_{d+1}^{\infty} {d^2(2k-d-1)(d+1)^{d \over d-1}\over k(k-1)}
  \left((d-1)k-{(d^2-d-2)}\right)^{1-2d \over d-1}\, dk
\end{eqnarray}

For $d=2$, equation (9) gives $C={46\over 3}- 36\ln{3\over 2}=0.7366 $, which is the same result obtained in~\cite{ZhYaZhFuWa04}.
For $d=3$ we obtain $C=18+36\sqrt{2}\arctan\sqrt{2}+
{9\over 2}\pi-18\sqrt{2}\,\pi = 0.8021$. For $d=4$ the analytical value is $0.8404$ and for $d=5$ it is $0.8658$. We note that they do not depend on the order of the network.
Note also that for $d=2$ the value obtained is not far from the value of the clustering in the coauthorship network in neuroscience ($0.76$) ~\cite{AlBa02}. Thus the clustering $C$ of HDRAN is high and increases with the dimension $d$ and approaches a limit of $1$ when $d$ is large.
In Fig. 3, we present simulation results for $C$ vs the dimension $d$ for various values of the $N$, the network order. 
For $d=2,3,4,5$, the simulation results  provide values of  $C$ of $0.768$, $0.825$, $0.855$ and  $0.877$, respectively.
We see that the small difference between analytical and simulation values decreases with $d$.
This difference might be explained because the continuum approach does not properly account all the vertices,  affecting $P(k)$ which is used in Eq. (9) to obtain $C$, see~\cite{DoMe01}. 

%%%%%%%%%%%%%%%%%%%%%%%%%%%%%%%%%%%%%%%%%%%%%%%%%%%%%%%%%%
% Figure  3
%%%%%%%%%%%%%%%%%%%%%%%%%%%%%%%%%%%%%%%%%%%%%%%%%%%%%%%%%%
\begin{figure}
\begin{center}
\includegraphics[width=8cm]{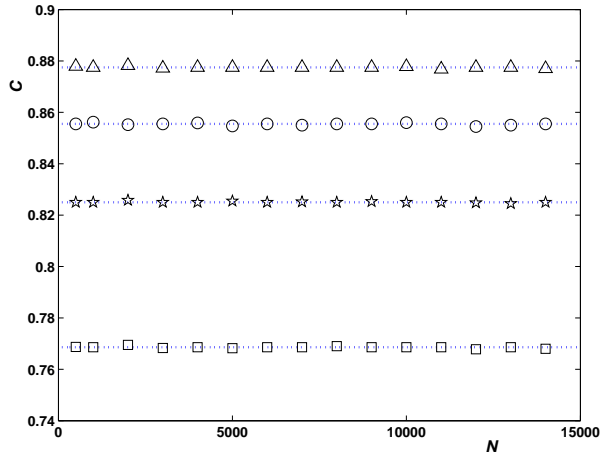}
\caption{Clustering coefficient $C$ of HDRAN at various values of
dimension $d$. The squares, stars, circles and triangles denote 
$d=2$, $d=3$, $d=4$ and $d=5$,
respectively. 
The dashed lines are fits to the simulated data.}
\label{fig:loglog}
\end{center}
\end{figure}
%%%%%%%%%%%%%%%%%%%%%%%%%%%%%%%%%%%%%%%%%%%%%%%%%%%%%%%%%%

\subsection{Average path length}
The average path length of a network (APL) is
defined as the number of edges in the shortest path between two vertices
averaged over all pairs of vertices.
Below, using the same approach than in ~\cite{ZhYaZhFuWa04}
we will discuss the APL of a high dimensional random Apollonian network.

First we note that it is not very difficult to prove that for a HDRAN and
for any two arbitrary vertices $i$ and $j$ each shortest path from
$i$ to $j$ does not pass through  any vertex $k$ satisfying that $k < \max(i, j)$.
%{\em Proof.} Using vertices' sequence
%$i\rightarrow x_1\rightarrow x_2\rightarrow \cdots x_n \rightarrow j$  to denote
%the shortest path from $i$ to $j$ of length $n+1$.
%Obviously, $n = 0$ is the trivial case. Suppose that $n>0$ and
%$x_k = \max(x_1, x_2,\cdots ,x_n)$ if $x_k < \max(i, j)$, then the proposition is true.
%
%In succession, we prove that the case $x_k > \max(i, j)$ would not emerge.
%Suppose that the vertex $x_k$ is created itself thanks to a $(d+1)$-clique
%$K$ composed of vertices $w_1,w_2, \cdots ,w_d,w_{d+1}$.
%Since $x_k= \max(x_1, x_2,\cdots ,x_n)$, according to the construction
%algorithm of HDRAN, of all the vertices younger than $x_k$ only $d+1$ vertices
%$w_1,w_2, \cdots ,w_d$, and $w_{d+1}$ are connected to $x_k$, then the path
%from $i$ to $j$ passing through $x_k$ must enter into and leave clique $K$.
%We may assume that the path enter into clique $K$ by vertex $w_1$ and leave
%from vertex $w_2$, then there exists a subpath of $P_{ij}$ from $w_1$ to $w_2$
%going through $x_k$, which is apparently longer than the direct path
%$w_1\rightarrow w_2$. Hence if $P_{ij}$ is the shortest path,
%the oldest vertex in the path must be either $i$ or $j$.
%$\Box$

  If $d(i,j)$ denotes the distance between $i$ and $j$, we introduce
the total distance of a HDRAN with order $N$ as $\sigma(N)$
%%%
%%% eq 10
\begin{equation}
\sigma(N) = \sum_{1\leq i< j \leq N}d(i,j)
\end{equation}
and we denote the APL by $L(N)$, defined as:
%%%
%%% eq 11
\begin{equation}
L(N) ={2\sigma(N)\over N(N-1)}
\end{equation}
According to the former remark, the addition of new vertices will not
affect the distance between those already existing and we have:
%%%
%%% eq 12
\begin{equation}
\sigma(N+1) = \sigma(N)+ \sum_{1}^{N}d(i,N+1)
\end{equation}
Assume that the vertex $N+1$ is added joining the $(d+1)$-clique
$K$ composed of vertices $w_1,w_2,\cdots ,w_d,w_{d+1}$ then Eq. (12) can be rewritten as:
%%%
%%% eq 13
\begin{equation}
\sigma(N+1) = \sigma(N)+ \sum_{1}^{N}(D(i,w)+1)=\sigma(N)+N+ \sum_{1}^{N}D(i,w)
\end{equation}
where $D(i,y) = \min\{d(i,w_1),d(i,w_2), \cdots ,d(i,w_d),d(i,w_{d+1}\}$.
Constricting the  clique $K$ continuously into a single vertex $w$
(here we assume that $w\equiv w_1$), we have $D(i,w)=d(i,w)$.
Since $d(w_1,w) = d(w_2,w) = \cdots=d(w_{d+1},w)= 0$, Eq. (13) can be rewritten as:
%%%
%%% eq 14
\begin{equation}
\sigma(N+1) =\sigma(N)+N+ \sum_{i\in \Gamma}d(i,w)
\end{equation}
where $\Gamma= \{1, 2,\cdots ,N\}-\{w_1,w_2,\cdots,w_d,w_{d+1}\}$ is a vertex
set with cardinality $N-d-1$.
The sum $\sum_{i\in \Gamma}d(i,w)$ can be considered as the total
distance from one vertex $w$ to all the other vertices in our model
with order $N-d$ and it can be
roughly approximated in terms of $L(N-d)$:
%%%
%%% eq 15
\begin{equation}
\sum_{i\in \Gamma}d(i,w)\approx (N-d-1) L(N-d)
\end{equation}
Note that, as $L(N)$ increases monotonously with $N$, it is clear that:
%%%
%%% eq 16
\begin{equation}
(N-d-1) L(N-d) = {2\sigma(N-d) \over N-d} < {2\sigma(N) \over N}
\end{equation}
From equations (14), (15) and (16), we can obtain the expression:
%%%
%%% eq 17
\begin{equation}
\sigma(N+1) < \sigma(N) +N + {2\sigma(N) \over N}
\end{equation}
Considering Eq. (17) as an equation and not an inequality, 
we can provide an upper bound for the variation of $\sigma(N)$ as
%%% eq 18
\begin{equation}
{d\sigma(N) \over dN} =  N + {2\sigma(N) \over N}
\end{equation}
Which gives
%%%
%%% eq 19
\begin{equation}
\sigma(N) = N^2\ln N + \alpha,
\end{equation}
where $\alpha$ is a constant. 
As $\sigma(N) \sim N^2\ln N $, we have $L(N) \sim \ln N$.
Note that as we have deduced equation (19) from
an inequality, then $L(N)$ increases at most as $\ln N$ with $N$.

In Fig. 4, we report the simulation results on the average
path length $L$ vs the  dimension $d$ for various values of the network
order $N$, which agree well with the analytic results.
Note that the APL decreases with the dimension $d$ as the networks become more dense.
The dependence of the APL on $d$, comes from the method used to obtain 
Eq. (14) as the sum there is on a set $\Gamma$ with cardinality  $N-d-1$.

%%%%%%%%%%%%%%%%%%%%%%%%%%%%%%%%%%%%%%%%%%%%%%%%%%%%%%%%%%
% Figure  4
%%%%%%%%%%%%%%%%%%%%%%%%%%%%%%%%%%%%%%%%%%%%%%%%%%%%%%%%%%
\begin{figure}
\begin{center}
\includegraphics[width=8cm]{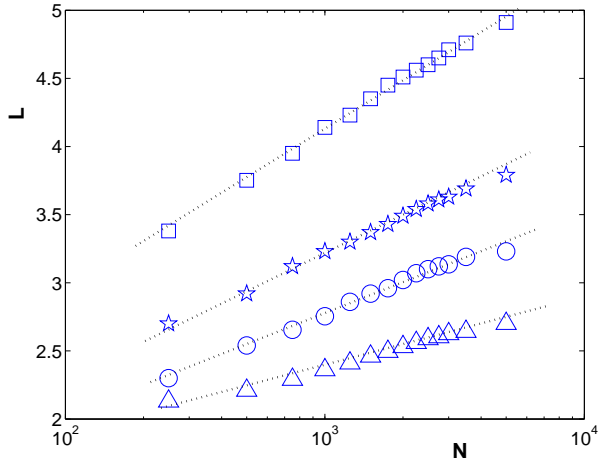}
\caption{Semilogarithmic plot of the  average path length $L$ of a HDRAN vs its network size $N$ for different values of the  dimension $d$. The squares, stars, circles and triangles  denote the simulation results for  $d=2$, $d=3$, $d=4$ and $d=5$, respectively. The lines are fits to the simulated data. One can see that $L$ grows at most as $\ln N$.}
\label{fig:cluster}
\end{center}
\end{figure}
%%%%%%%%%%%%%%%%%%%%%%%%%%%%%%%%%%%%%%%%%%%%%%%%%%%%%%%%%%
\section{Conclusion and discussion}
In summary, we have introduced an iterative algorithm for the construction of
high-dimensional random Apollonian networks associated with high
dimensional random packings. The networks have the typical
characteristics of many technological and social networks and are
small-world  with power-law degree distributions, thus they
can model a variety of scale-free networks of the real-life world.
We have computed the analytical expressions for the degree distribution and
clustering coefficient and we show that they depend on the dimension.
Also, the average path length of the networks grows logarithmically with
the order of the HDRAN. Moreover, HDRAN are a generalization of
the maximal planar network introduced in \cite{ZhYaWa04}.
Future work should include studying in detail processes such as percolation,
spreading, searching and diffusion taking place on HDRANs.
\subsection*{Acknowledgment}
This research was supported by the National Natural Science
Foundation of China under Grant No. 70431001.
Support for F.C. was provided by the Secretaria de Estado de Universidades
e Investigaci\'on (Ministerio de Educaci\'on y Ciencia),  Spain, and the
European Regional Development Fund (ERDF) under project TIC2002-00155.
The authors are grateful to the anonymous referees
for their valuable comments and suggestions.
%%%%%%%%%%%%%%%%%%%%%%%%%%%%%%%%%%%%%%%%%%%%%%%%%%%%%%%%%%%%%%%%%
%%%%%%%%%%%%%%%%%%%%%%%%%%%%%%%%%%%%%%%%%%%%%%%%%%%%%%%%%%%%%%%%%


\begin{thebibliography}{10}

%1
\bibitem{WaSt98}
D.J. Watts, S.H. Strogatz,
Collective dynamics of `small-world' networks,
Nature 393 (1998) 440--442.

%2
\bibitem{BaAl99}
A.-L. Barab\'asi, R. Albert,
Emergence of scaling in random networks,
Science 286 (1999) 509--512.


%% 3
\bibitem{AlBa02}
R. Albert, A.-L. Barab\'asi,
Statistical mechanics of complex networks,
Rev. Mod. Phys.  74  (2002)   47--97.

%% 4
\bibitem{DoMe02}
S.N. Dorogvtsev, J.F.F. Mendes,
Evolution of networks,
Adv. Phys. 51 (2002) 1079--1187.

%% 5
\bibitem{Ne03}
M.E.J. Newman,
The structure and function of complex networks,
SIAM Review 45 (2003) 167--256.

%% 6
\bibitem{FaFaFa99}
M. Faloutsos, P. Faloutsos, C. Faloutsos,
On power-law relationships of the internet topology,
Comput. Commun. Rev. 29 (1999) pp. 251--260.

%% 7
\bibitem{AlJeBa99}
R. Albert, H. Jeong,  A.-L. Barab\'asi,
Diameter of the world wide web,
Nature  401  (1999) 130--131.

%% 8
\bibitem{JeToAlOlBa00}
H. Jeong, B. Tombor, R. Albert, Z.N. Oltvai, A.-L. Barab\'asi,
The large-scale organization of metabolic networks,
Nature 407 (2000) 651--654.

%% 9
\bibitem{JeMaBaOl01}
H. Jeong, S. Mason, A.-L. Barab\'asi, Z.N. Oltvai,
Lethality and centrality in protein networks,
Nature 411 (2001) 41--42.

%% 10
\bibitem{Ne01}
M.E.J. Newman,
The structure of scientific collaboration networks,
Proc. Natl. Acad. Sci. U.S.A. 98 (2001) 404--409.

%% 11
\bibitem{LiEdAmStAb01}
F. Liljeros, C.R. Edling, L.A.N. Amaral, H.E. Stanley, Y. \AA{berg},
The web of human sexual contacts,
Nature 411 (2001) 907--908.

%% 12
\bibitem{BaRaVi01}
A.-L. Barab\'asi, E. Ravasz,T. Vicsek,
Deterministic scale-free networks,
Physica A   299  (2001) 559--564.


%% 13
\bibitem{DoMeSa00}
S.N. Dorogovtsev, J.F.F. Mendes, A.N. Samukhin,
Structure of growing networks with preferential linking,
Phys. Rev. Lett.  85 (2000) 4633--4636.

%% 14
\bibitem{KaReLe00}
P.L. Krapivsky, S. Redner,  F. Leyvraz,
Connectivity of growing random networks,
Phys. Rev. Lett. 85 (2000) 4629--4632.

%% 15
\bibitem{AmScBaSt00}
L. A. N., Amaral, A. Scala, M. Barth\'el\'emy,  H. E. Stanley,
Classes of small-world networks,
Proc. Natl. Acad. Sci. U.S.A.  97 (2000) 11149.

%% 16
\bibitem{DoMe00}
S. N. Dorogovtsev, J. F. F. Mendes,
Evolution of networks with aging of sites,
Phys. Rev. E 62 (2000) 1842.

%% 17
\bibitem{BiBa01}
G. Bianconi,  A.-L. Barab\'asi,
Competition and multiscaling in evolving networks,
Europhys. Lett. 54 (2001) 436--442.

%% 18
\bibitem{AlBa00}
R.Albert,  A.-L. Barab\'asi,
Topology of evolving networks: Local events and universality,
Phys. Rev. Lett. 85 (2000) 5234--5237.

%% 19
\bibitem{DoMe00b}
S. N.Dorogovtsev,  J. F. F. Mendes,
Scaling behaviour of developing and decaying networks,
Europhys. Lett. 52 (2000) 33--39.

%% 20
\bibitem{KlKuRaRaTo99}
J. M.Kleinberg, R. Kumar, P. Raghavan, S. Rajagopalan,  A. Tomkins,
The Web as a graph: Measurements, models and methods,
Proceedings of the 5th Annual International Conference, COCOON'99,
Tokyo, July 1999 (Springer-Verlag, Berlin), (1999) 1.

%% 21
\bibitem{KuRaRaSiToUp00}
R. Kumar, P. Raghavan, S. Rajalopagan, D. Sivakumar, A. S.
Tomkins, E. Upfal, 
The Web as a graph, Proceedings of the 19th
Symposiumon Principles of Database Systems (2000)  1.

%% 22
\bibitem{KuRaRaSiToUp00b}
R. Kumar, P. Raghavan, S. Rajalopagan, D. Sivakumar, A. S.
Tomkins,  E. Upfal, 
Stochastic models for the Web graph,
Proceedings of the 41st IEEE Symposium on Foundations of Computer
Science (IEEE Computing Society, Los Alamitos, Calif.)
(2000) 57--65.

%% 23
\bibitem{ChLuDeGa03}
F. Chung, Linyuan Lu, T. G. Dewey,  D. J. Galas,
Duplication models for biological networks,
J. of Comput. Biology  10 (5) (2003) 677--688.

%% 24
\bibitem{KrRe01}
P. L.Krapivsky, S. Redner,
Organization of growing random networks,
Phys. Rev. E 63 (2001) 066123.

%% 25
\bibitem{Va01}
A. V\'azquez, 
Disordered networks generated by recursive searches,
Europhys. Lett. 54 (2001) 430--435.

%% 26
\bibitem{CoFeRa04}
F. Comellas, G. Fertin, A. Raspaud,
Recursive graphs with small-world scale-free properties,
Phys. Rev. E   69 (2004) 037104.


%% 27
\bibitem{CoSa02}
F. Comellas, M. Sampels,
Deterministic small-world networks,
Physica A  309 (2002) 231--235.

%% 28
\bibitem{DoGoMe02}
S.N. Dorogovtsev, A.V. Goltsev, J.F.F. Mendes,
Pseudofractal scale-free web,
Phys. Rev. E   65  (2002) 066122.

%% 29
\bibitem{RaBa03}
E. Ravasz, A.-L. Barab\'asi,
Hierarchical organization in complex networks,
Phys. Rev. E 67 (2003) 026112.


%% 30
\bibitem{No03}
J.D. Noh,
Exact scaling properties of a hierarchical network model,
Phys. Rev. E   67  (2003) 045103.

%% 31
\bibitem{JuKiKa02}
S. Jung, S. Kim, B. Kahng,
Geometric fractal growth model for scale-free networks,
Phys. Rev. E 65 (2002) 056101.

%% 32
\bibitem{ZhWaHuCh04}
T. Zhou, B.H. Wang, P.M. Hui,  K.P. Chan,
Integer networks,
arXiv: cond-mat/0405258.

%% 33
\bibitem{AnHeAnSi05}
J. S. Andrade Jr., H. J. Herrmann, R. F. S. Andrade   L. R. da Silva,
Apollonian Networks: Simultaneously scale-free, small world, Euclidean, space filling,
and with matching graphs,
Phys. Rev. Lett. 94 (2005) 018702.

%% 34
\bibitem{DoMa05}
J. P. K. Doye, C. P. Massen.
Self-similar disk packings as model spatial scale-free networks,
Phys. Rev. E 71 (2005) 016128.

%% 35
\bibitem{ZhCoFeRo05}
Z.Z. Zhang, F. Comellas, G. Fertin, L.L. Rong, 
High dimensional Apollonian networks,
arXiv:  cond-mat/0503316.

%% 36
\bibitem{ZhYaZhFuWa04}
T. Zhou, G. Yan, P. L. Zhou, Z. Q. Fu,   B. H Wang,
Random Apollonian networks,
arXiv:  cond-mat/0409414.

%% 37
\bibitem{So36}
F. Soddy,
The Kiss Precise,
Nature 137 (1936) 1021.


%% 38
\bibitem{Go37}
T. Gosset, %Thorold Gosset
The Kiss Precise,
Nature 139 (1937)  62. %(9 January 1937)

%% 
%\bibitem{Go37}
%T. Gosset. %Thorold Gosset
%The Hexlet. 
%Nature 139 (1937) 251. %251-252


%% 39
\bibitem{ZhYaWa04}
T. Zhou, G. Yan, B. H Wang,
Maximal planar networks wiht large clustering coefficient and power-law
degree distribution,
Phys. Rev. E 71 (2005) 046141.



%% 40
\bibitem{We01}
D.B. West,
{\em Introduction to Graph Theory}.
(Prentice-Hall,  Upper Saddle River, NJ, 2001).
% ISBN 0-13-014400-2

%% 41
\bibitem{BaAlJe99}
A.-L. Barab\'asi, R. Albert, H. Jeong,
Mean-field theory for scale-free random networks,
Physica A   272  (199) 173--187.

%% 42
\bibitem{DoMe01}
S. N. Dorogovtsev,   J. F. F. Mendes,
Comment on ``Breakdown of the internet under intentional attack'',
Phys. Rev. Lett. 87 (2001) 219801. 

\end{thebibliography}
\end{document}